\documentclass[11pt]{article}

\usepackage[english]{babel}

\usepackage{graphicx}
\usepackage[colorlinks=true, allcolors=blue]{hyperref}

\usepackage[title]{appendix}

\usepackage{amssymb,amsmath,amsthm,amsfonts,amstext}
\usepackage[left=1.5cm,right=1.5cm,top=1.5cm,bottom=1.5cm]{geometry}
\usepackage{enumerate}

\usepackage[mathscr]{euscript}

\usepackage{soul}

\usepackage[dvipsnames]{xcolor}
\usepackage{tabularx}

\usepackage{tikz}
\usetikzlibrary{arrows}
\usetikzlibrary{arrows.meta}
\usetikzlibrary{shapes}
\usetikzlibrary{backgrounds}
\usetikzlibrary{positioning}
\usetikzlibrary{decorations.markings}
\usetikzlibrary{patterns}
\usetikzlibrary{calc}
\usetikzlibrary{fit}
\usetikzlibrary{decorations}
\usetikzlibrary{decorations.pathreplacing}

\usepackage[framemethod=TikZ]{mdframed}

\usepackage[noend]{algpseudocode}
\makeatletter
\def\BState{\State\hskip-\ALG@thistlm}
\makeatother

\usepackage[ruled]{algorithm2e}

\newtheorem{mdalgorithm}{Algorithm}

\newenvironment{ourbox}{\begin{mdframed}[hidealllines=false,innerleftmargin=10pt,backgroundcolor=white!10,innertopmargin=2pt,innerbottommargin=5pt,roundcorner=10pt]}{\end{mdframed}}

\newtheorem{theorem}{Theorem}

\newtheorem{proposition}[theorem]{Proposition}
\newtheorem{lemma}[theorem]{Lemma}
\newtheorem{fact}{Fact}

\newtheorem{example}{Example}

\newcommand\IGNORE[1]{}

\newcommand{\Q}{\ensuremath{\mathbb Q}}

\newcommand{\safe}{\mathscr{S}}

\newcommand{\F}{\mathcal{F}}

\newcommand{\uv}[1]{\vec{v}_{\{#1\}}}

\title{Symmetric Submodular Functions, Uncrossable Functions, and Structural Submodularity}

\author{
\large
Miles Simmons\thanks{
        {\tt mjsimmons@uwaterloo.ca}.
        Department of Combinatorics \& Optimization, University of Waterloo, Canada.}
\and
Ishan Bansal\thanks{
        {\tt ib332@cornell.edu}.
	Amazon, Bellevue, WA, USA. This work is external and does not relate to the position at Amazon. }
\and
Joseph Cheriyan\thanks{
{\tt jcheriyan@uwaterloo.ca}.
        Department of Combinatorics \& Optimization, University of Waterloo, Canada.}
}

\begin{document}
\maketitle

\begin{abstract}{
Diestel~et~al.\ \cite{DEW2019} introduced the notion of abstract separation systems
that satisfy a submodularity property, and they call this {structural submodularity}.

Williamson~et~al.\ \cite{WGMV95} call a family of sets $\F$ {uncrossable} if the following holds:
for any pair of sets $A,B\in\F$,
both $A\cap{B},A\cup{B}$ are in $\F$, or both $A-B,B-A$ are in $\F$.
Bansal~et~al.\ \cite{BCGI24} call a family of sets $\F$ {pliable} if the following holds:
for any pair of sets $A,B\in\F$,
at least two of the sets $A\cap{B},A\cup{B},A-B,B-A$ are in $\F$.
We say that a pliable family of sets $\F$ satisfies structural submodularity if the following holds:
for any pair of crossing sets $A,B\in\F$,
at least one of the sets $A\cap{B},A\cup{B}$ is in $\F$, and
at least one of the sets $A-B,B-A$ is in $\F$.

For any positive integer $d\geq2$,
we construct a pliable family of sets $\F$ that satisfies structural submodularity such that
(a)~there do not exist a symmetric submodular function $g$ and $\lambda\in\Q$ such that
\hbox{$\F = \{ S \,:\, g(S)<\lambda \}$}, and
(b)~$\F$ cannot be partitioned into $d$ (or fewer) uncrossable families.
}
\end{abstract}

\section{Introduction \label{sec:intro}}
{
Diestel~et~al.\ \cite{DEW2019,D2017,DEE2017,DO2019,EKT2021} introduced the notion of
abstract separation systems that satisfy a submodularity property, and
they call this structural submodularity.
One of their motivations was to
identify the few structural assumptions one has to make of a set
of objects called ‘separations’ in order to capture the essence of
tangles in graphs, and thereby make them applicable in wider contexts.

Decades earlier, Williamson~et~al.\ \cite{WGMV95} defined a family of sets $\F$ to be \textit{uncrossable} if
the following holds:
for any pair of sets $A,B\in\F$,
both $A\cap{B},A\cup{B}$ are in $\F$, or both $A-B,B-A$ are in $\F$.
They used this notion to design and analyse a primal-dual approximation algorithm for covering an
uncrossable family of sets, and they proved an approximation guarantee of two for their algorithm.
Recently, Bansal~et~al.\ \cite{BCGI24} defined a family of sets $\F$ to be \textit{pliable} if
the following holds:
for any pair of sets $A,B\in\F$,
at least two of the (four) sets $A\cap{B},A\cup{B},A-B,B-A$ are in $\F$.
Bansal~et~al.\ \cite{BCGI24} showed that the primal-dual algorithm of Williamson~et~al.\ \cite{WGMV95}
achieves an approximation guarantee of $O(1)$ for
the problem of covering a pliable family of sets that satisfies property~$(\gamma)$.
(We discuss property~$(\gamma)$ in the following section;
it is a combinatorial property, and the analysis of \cite{BCGI24} relies on it,
but it is not relevant for this paper.)
Simmons, in his thesis, \cite{S2025}, uses the notion of a strongly pliable family of sets.
This notion is the same as the notion of structural submodularity of Diestel~et~al.\ \cite{DEW2019},
and, in this paper, we use the term structural submodularity (rather than strongly pliable).
We say that a pliable family of sets $\F$ satisfies \textit{structural submodularity} if the following holds:
for any pair of crossing sets $A,B\in\F$,
at least one of the sets $A\cap{B},A\cup{B}$ is in $\F$, and
at least one of the sets $A-B,B-A$ is in $\F$.

A natural way to obtain a pliable family of sets $\F$ that satisfies {structural submodularity}
is to take the ``sublevel~sets'' of any symmetric submodular function, that is, pick
\hbox{$\F = \{ S\subseteq{V} : g(S) < \lambda \}$}, where
$g:2^V\rightarrow\Q$ is a symmetric submodular function and $\lambda\in\Q$.
This raises the question whether every pliable family that satisfies {structural submodularity}
corresponds to the ``sublevel~sets'' of a symmetric submodular function.
We answer this question in the negative by constructing a particular pliable family $\F$ that
satisfies {structural submodularity} such that there exist no symmetric submodular function $g$ and
$\lambda\in\Q$ such that \hbox{$\F = \{ S : g(S) < \lambda \}$} (Proposition~\ref{pro:no-submod-realization}).
Moreover, given any positive integer $d\geq2$, our construction ensures that $\F$ cannot be
partitioned into $d$ (or fewer) uncrossable families (Proposition~\ref{pro:partition-uncrossable}).

The results in this paper are based on a sub-chapter of the first author's thesis,
see \cite[Chapter~2.3.2]{S2025}.

\bigskip
\bigskip

\hbox{
\begin{minipage}{\textwidth}
{
\begin{example}
The following example shows a pliable family $\F$ that satisfies structural submodularity such that
there exist no symmetric submodular function $g$ and $\lambda\in\Q$ such that
\hbox{$\F = \{ S : g(S) < \lambda \}$}.
See section~\ref{sec:construct} for more details.
Let $V$ be the set of binary vectors of length~3.
For notational convenience, we label the vectors in $V$ by the digits $0,\dots,7$
such that $0=\vec{v}_{\{\}}$, $1=\vec{v}_{\{1\}}$, $2=\vec{v}_{\{2\}}$, $3=\vec{v}_{\{1,2\}}$,
$4=\vec{v}_{\{3\}}$, $5=\vec{v}_{\{1,3\}}$, $6=\vec{v}_{\{2,3\}}$, $7=\vec{v}_{\{1,2,3\}}$.
\begin{align*}
\F = \{	& V_1 = \{1,3,5,7\}, V_2 = \{2,3,6,7\}, V_3 = \{4,5,6,7\}, \\
	& \{2, 6\}, \{4,5\}, \{4, 6\}, \{3, 7\}, \{5, 7\}, \{6, 7\}
	\{3\}, \{4\}, \{5\}, \{6\}, \{7\}, \{4, 5, 7\} \}
\end{align*}
Let $U_3 = V_3 - (V_1-V_2)$, and let $W_3 = U_3 - (V_2-V_1)$;
note that $(V_1-V_2-V_3) = \{{1}\}$, and $(V_2-V_1-V_3) = \{{2}\}$.
We write the submodular inequalities for $3$ pairs of crossing sets, then we sum the $3$ inequalities:
\begin{align*}
V_1,& \quad V_2 		& g(V_1) + g(V_2) - g(V_1-V_2) - g(V_2-V_1) &\geq 0 \\
(V_1-V_2),& \quad V_3		& g(V_1-V_2) + g(V_3) - g(V_1-V_2-V_3) - g(U_3) &\geq 0 \\
(V_2-V_1),& \quad U_3		& g(V_2-V_1) + g(U_3) - g(V_2-V_1-V_3) - g(W_3) &\geq 0 \\
\text{Sum of inequalities:}	&&
	g(V_1) + g(V_2) + g(V_3) - g(\{{1}\}) - g(\{{2}\}) - g(W_3) \geq 0
\end{align*}
Since $V_1,V_2,V_3\in\F$, we have $g(V_1)<\lambda, g(V_2)<\lambda, g(V_3)<\lambda,$ and
since $\{{1}\}, \{{2}\}, W_3=\{4,7\}\notin\F$ we have
$g(\{{1}\})\geq\lambda, g(\{{2}\})\geq\lambda, g(W_3)\geq\lambda$.
Contradiction.
\end{example}
}
\end{minipage}
}

}
\section{	Preliminaries \label{sec:prelims}}
{
For a positive integer $k$, we use $[k]$ to denote the set $\{1,2,\dots,k\}$.
A pair of subsets $A,B$ of $V$ (the ground-set) is said to \textit{cross} if each of the four sets
$A\cap{B}, V-(A\cup{B}), A-B, B-A$ is non-empty.

A function $g:2^V\rightarrow\Q$ on subsets of $V$ is called \textit{submodular} if
the inequality $g(A) + g(B) \geq g(A \cap B) + g(A \cup B)$ holds
for all pairs of sets $A,B \subseteq V$, \cite{Schrijver}.
A function $g:2^V\rightarrow \mathbb{Q}$ is called \textit{symmetric} if
$g(S) = g(\bar{S}) = g(V-S)$, for all sets $S\subseteq{V}$.
For a symmetric submodular function $g:2^V\rightarrow \mathbb{Q}$, we have
\[ g(A) + g(B) = g(A) + g(\bar{B}) \geq g(A \cap \bar{B}) + g(A \cup \bar{B}) = g(A-B) + g(B-A), \]
since $A\cap\bar{B} = A-B$ and
$g(A\cup\bar{B}) = g(\overline{A\cup\bar{B}}) = g(\bar{A}\cap{B}) = g(B-A)$.

Diestel~et~al.\ \cite{DEW2019} call a subset $M$ of a lattice $(L,\vee,\wedge)$
\textit{submodular} if for all $x,y\in{M}$ at least one of $x\vee{y}$ and $x\wedge{y}$ lies in $M$.
A \textit{separation system} $(\vec{\safe}, \leq, *)$ is a partially ordered set with an
order-reversing involution $*$.
The elements of $\vec{\safe}$ are called oriented separations.
A separation system $\vec{\safe}$ contained in a given universe $\vec{U}$ of separations
is \textit{structurally submodular} if it is submodular as a subset of the lattice underlying $\vec{U}$.

Next, we discuss property~$(\gamma)$ for a family of sets $\F$,
though this property is not used in this paper.
A family of sets $\F$ satisfies property~$(\gamma)$ if
for any sets $C, S_1, S_2 \in\F$ such that
$S_1\subsetneq S_2$, $C$ is inclusion-wise minimal, and $C$ crosses both $S_1,S_2$,
the set $S_2 - (S_1\cup C)$ is either empty or is in $\F$, \cite{BCGI24}.
}
\section{	Construction of family of sets $\F$ \label{sec:construct}}
{

Let $k\ge3$ be a positive integer.
Let $V$ be the set of binary vectors of length $k$.
We denote an elements of $V$ by $\vec{v}$ and we denote the coordinates of $\vec{v}$ by
$\vec{v}_1,\vec{v}_2,\dots,\vec{v}_k$.
For a set of indices $I\subseteq[k]$, we use $\vec{v}_I$ to denote $\vec{v}\in{V}$ such that
$\vec{v}_j=1$ iff $j\in{I}$.
For example, $\vec{v}_{\{1\}} = (1,0,\dots,0)$, $\vec{v}_{\{1,k\}} = (1,0,\dots,0,1)$, and
$\vec{v}_{[k]}$ is the vector with a one in each coordinate.
Let us call $\vec{v}_{\{1\}},\vec{v}_{\{2\}},\dots,\vec{v}_{\{k\}}$ the \textit{unit-vectors}.
For an index $i\in[k]$, let $V_i = \{ \vec{v}\in V \;:\; \vec{v}_i=1 \}$;
thus, $V_i$ is the set of vectors in $V$ that have a one in the $i$-th coordinate.

Observe that a unit-vector is in exactly one of the sets $V_1,\dots,V_k$, e.g.,
$\vec{v}_{\{1\}}$ is in $V_1$ and it is in none of $V_2,\dots,V_k$.
Moreover, observe that the sets $V_i$ and $V_j$ cross, for any $i,j\in[k]$ such that $i\not=j$.

Algorithm~\ref{alg:csccounter} constructs the required family $\F$.

\begin{algorithm}\caption{Family $\F$ Construction}\label{alg:csccounter}
\vspace{5pt}
\textbf{Initialize:} $\F = \F_0 = \{V_1,...,V_k\}, \ell = 1$.
        
        \vspace{10pt} 
         Begin iteration $\ell$, let $\F_\ell = \emptyset$:
        \begin{enumerate}
            \item Examine every pair of sets $A, B$ in $\F$ that cross.
            \begin{enumerate}
                \item If $A \cap B \notin \F$, add $A \cap B$ to $\F_\ell$.
            \item If $A - B, B - A \notin \F$,
            \begin{enumerate}
                \item If both $A - B$ and $B - A$ contain unit-vectors, then add to $\F_{\ell}$ the set containing the unit-vector of larger index (i.e., suppose $A-B\owns\vec{v}_i, B-A\owns\vec{v}_j, j>i$, then add $B-A$ to $\F_{\ell}$).
                \item If one of $A - B$ or $B - A$ contains a unit-vector and the other contains no unit-vector, then add the latter set (containing no unit-vector) to $\F_\ell$.
                \item Otherwise, add one of $A-B$ or $B-A$ to $\F_\ell$ (arbitrary choice).
            \end{enumerate}
            \end{enumerate}
        \item If $\F_\ell = \emptyset$, all pairs of crossing sets have the required subsets in $\F$.  Return $\F$, a family that satisfies structural submodularity.
        \item Otherwise, add all sets in $\F_\ell$ to $\F$, update $\ell \rightarrow \ell+1$, and proceed to the next iteration.
    \end{enumerate}
\end{algorithm}

Figure~\ref{fig:csccounteriterations} illustrates the iterative construction of $\F$.

{
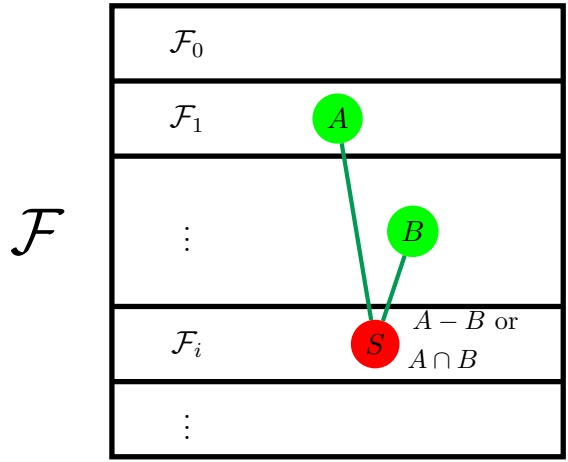
\begin{figure}[hbt]
    \centering
\begin{tikzpicture}
   \draw[black, line width=2pt]
      (0,0) rectangle (6,6);
       \draw[black, line width=2pt]
      (0,1) -- (6,1);
       \draw[black, line width=2pt]
      (0,2) -- (6,2);
      \draw[black, line width=2pt]
      (0,4) -- (6,4);
      \draw[black, line width=2pt]
      (0,5) -- (6,5);
      \node at (1,5.5) {\large{$\mathcal{F}_0$}};
      \node at (1,4.5) {\large{$\mathcal{F}_1$}};
      \node at (1,1.5) {\large{$\mathcal{F}_i$}};
      \node at (-1,3) {\Huge{$\mathcal{F}$}};
      \node at (1,3) {\large{$\vdots$}};
      \node at (1,0.5) {\large{$\vdots$}};

    \node at (4.7,1.8) {\small{$A-B$ or}};
    \node at (4.4,1.3) {\small{$A \cap B$}};

    \tikzstyle{every node}=[circle,semithick,minimum size=2pt,outer sep=0 pt,inner sep = 2.8pt];

    \node[fill = green] (A) at (3,4.5) {$A$};

    \node[fill = green] (B) at (4,3) {$B$};

    \node[fill = red] (S) at (3.5,1.5) {$S$};

    \draw[ForestGreen,ultra thick] (A) edge (S);
    \draw[ForestGreen,ultra thick] (B) edge (S);
      
\end{tikzpicture}
    \caption{The iterative construction of the family $\F$ returned by
Algorithm~\ref{alg:csccounter} is illustrated; $\F_0,\F_1,\dots,\F_i$ denote the sub-family of sets
added to $\F$ in iteration $0,1,\dots,i$. In iteration $i$, for each crossing pair of sets $A,B$ in
the current $\F$, some of the sets $A\cap{B},A-B,B-A$ (not in the current $\F$) are placed in $\F_i$.}
    \label{fig:csccounteriterations}
\end{figure}
}

\clearpage

{
\hbox{
\begin{minipage}{\textwidth}
{
\begin{example}
We present an example for $k = 3$.  The names of the nodes are displayed in Figure~\ref{fig:csccounter}.

For notational convenience, we label the vectors in $V$ by the digits $0,\dots,7$
such that $0=\vec{v}_{\{\}}$, $1=\vec{v}_{\{1\}}$, $2=\vec{v}_{\{2\}}$, $3=\vec{v}_{\{1,2\}}$,
$4=\vec{v}_{\{3\}}$, $5=\vec{v}_{\{1,3\}}$, $6=\vec{v}_{\{2,3\}}$, $7=\vec{v}_{\{1,2,3\}}$.

$\mathcal{F}_0 = \{V_1 = \{1,3,5,7\}, V_2 = \{2,3,6,7\}, V_3 = \{4,5,6,7\}\}$

$\mathcal{F}_1 = \{ \{2, 6\}, \{4,5\}, \{4, 6\}, \{3, 7\}, \{5, 7\}, \{6, 7\} \}$

$\mathcal{F}_2 = \{ \{3\}, \{4\}, \{5\}, \{6\}, \{7\}, \{4, 5, 7\} \}$

$\mathcal{F} = \mathcal{F}_0 \cup \mathcal{F}_1 \cup \mathcal{F}_2$ satisfies structural submodularity.
\end{example}
}
\end{minipage}
}

{
\begin{figure}[hbt]
    \centering
    \begin{tikzpicture}
        \draw[ForestGreen, line width=2pt,radius = 52.5pt] (0,0) circle;
        \node at (-1,2) {$V_1$};
        \draw[blue, line width=2pt,radius = 52.5pt] (2,0) circle;
        \node at (3,2) {$V_2$};
        \draw[purple, line width=2pt,radius = 60pt] (1,-2) circle;
        \node at (3.25,-3) {$V_3$};

           \tikzstyle{every node}=[circle,fill=black,semithick,minimum size=2pt,outer sep=0 pt,inner sep = 2.8pt];

        \node[label=above:$\vec{v}_{\{1\}}$] at (-1,0) {};
        \node[label=above:$\vec{v}_{\{2\}}$] at (3,0) {};
        \node[label={[label distance=-0.5em]90: $\vec{v}_{\{3\}}$}] at (1,-3) {};
        \node[label={[label distance=-0.85em]90: $\vec{v}_{\{1,3\}}$}] at (0,-1.5) {};
        \node[label={[label distance=-0.85em]75: $\vec{v}_{\{2,3\}}$}] at (2,-1.5) {};
        \node[label={[label distance=-1em]90: $\vec{v}_{\{1,2\}}$}] at (1,0.5) {};
        \node[label={[label distance=-1em]90: $\vec{v}_{\{1,2,3\}}$}] at (1,-1) {};
        \node[label={[label distance=-0.5em]90: $\vec{v}_{\{\}}$}] at (4,-2) {};
    \end{tikzpicture}
     \caption{Illustration of Algorithm~\ref{alg:csccounter} for $k = 3$.
	The vectors $\vec{v}\in{V}$ and the sets $V_1,V_2,V_3$ are illustrated.}
     \label{fig:csccounter}
     
 \end{figure}
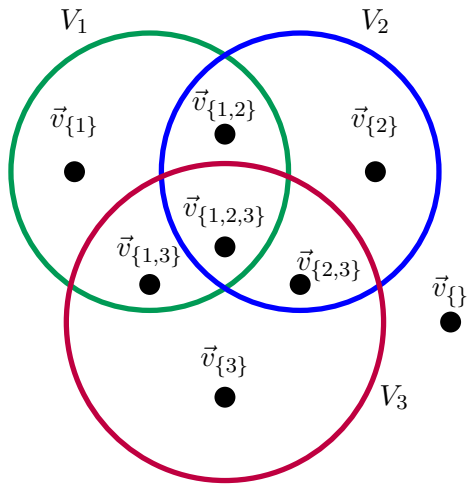
}

}

\section{Analysis of family of sets $\F$ \label{sec:analysis}}

Our analysis of the algorithm relies on several lemmas.
The first lemma focuses on the sets $V_1,\dots,V_k$ that are placed in $\F$ at the start.
The second lemma states that Algorithm~\ref{alg:csccounter} terminates.

The third lemma, Lemma~\ref{lem:keyclaim}, is our key lemma.
This lemma allows us to write a set $S\in\F$ containing a unit-vector say $\vec{v}_i$
(such that $S\notin\{V_1,\dots,V_k\}$) as the difference of two crossing sets $S',S''$ of $\F$ that
each contain a unit-vector such that $S''$ is from an earlier iteration than $S$
and the index of the unit-vector in $S''$ is smaller than $i$;
moreover, $S'$ is from an earlier iteration than $S$ or from the same iteration as $S$.
Based on this lemma, we derive relevant properties of the family $\F$ in
Lemmas~\ref{lem:rewriteSinF}, \ref{lem:propertySinF}, \ref{lem:largecuts}.

This section concludes with Proposition~\ref{pro:partition-uncrossable} which shows that $\F$ cannot
be partitioned into $d<k$ uncrossable families.

\begin{lemma}\label{lem:subsets}
Let $\mathcal{F}$ be the output of Algorithm~\ref{alg:csccounter}.

\begin{itemize}
\item[(i)]
Each set $S\in\F$ is a subset of one of the sets $V_1,\dots,V_k$.

\item[(ii)]
Each set $S\in\F$ contains at most one unit-vector;
moreover, if $S$ contains $\uv{i}$, $i\in[k]$, then $S$ is a subset of $V_i$.
\end{itemize}
\end{lemma}

\begin{proof}
(i)~By induction on the index $i$ of the sub-family $\F_i$ that contains $S$.
The induction hypothesis states that each set in $\F_0,\dots,\F_{i-1}$ is a subset of
one of the sets $V_1,\dots,V_k$.
The induction basis holds since $S\in\F_0$ implies that $S=V_i$ for some $i\in[k]$.
For the induction step, observe that $S\in\F_i$ implies that $S=A\cap{B}$ or $S=A-B$,
for sets $A,B\in\F$ that were added to $\F$ in an earlier iteration.
By the induction hypothesis, $A \subseteq V_j$ for some $j \in [k]$. Hence, $A \cap B, A - B \subseteq V_j$.

(ii)~The second part follows from the first part and the definition of the sets $V_1,\dots,V_k$.
\end{proof}

\begin{lemma} Algorithm~\ref{alg:csccounter} terminates.
\end{lemma}

\begin{proof}
By Lemma~\ref{lem:subsets}, every set added to $\F$ by
the algorithm is a subset of one of the sets $V_1,\dots,V_k$.  There
are a finite number of these subsets.  Clearly, the family consisting
of all subsets of the sets $V_1,\dots,V_k$ satisfies structural submodularity.
\end{proof}

\begin{fact}\label{fact:crossing}
Let $A,B\in\F$ be sets that each contain a unit-vector.
If the unit-vectors in $A,B$ are distinct, then $A,B$ cross iff $A\cap{B}$ is non-empty.
\end{fact}

\begin{proof}
$A$ and $B$ cross if $A \cap B, V - (A \cup B), A-B, B-A$ are all non-empty.
Suppose $\uv{i}$ is in $A$, $\uv{j}$ is in $B$, and $i \neq j$.
Since $\uv{i} \in A, \uv{i} \notin B$, and $\uv{j} \in B, \uv{j} \notin A$ (by Lemma~\ref{lem:subsets}),
we have $A-B, B-A \neq \emptyset$.
Also, note that $\vec{v}_{\{\}} \in V - (A \cup B)$, since $\vec{v}_{\{\}} \notin V_1\cup\dots\cup{V_k}$.
Thus, $A,B$ cross iff $A\cap{B}$ is non-empty.
\end{proof}

\begin{lemma} \label{lem:keyclaim}
Let $S\in\F_{\ell}, \ell\geq1$ be a set that contains a unit-vector, say $\vec{v}_i \in S$, $i\in[k]$.
Then there is a crossing pair of sets $S', S'' \in\F$ such that $S = S'-S''$, and we have
$S'\owns\vec{v}_i$, $S'\in  \bigcup_{h=0}^{\ell}\mathcal{F}_{h}$,
$S''\owns\vec{v}_j$, $j<i$, $S''\in \bigcup_{h=0}^{\ell-1}\mathcal{F}_{h}$.
\end{lemma}

\begin{proof}
By induction on the index $\ell$ of the sub-family $\F_{\ell}$ that contains $S$.

\vspace{-2ex}

\begin{description}
\item{Induction Hypothesis:}
Let $S \in \F_{\ell}$ be a set that contains a unit-vector, say $\vec{v}_{\{i\}}\in{S}$.
Then there exists a pair of crossing sets
$S'\in \bigcup_{h=0}^{\ell}\mathcal{F}_{h}$,
$S''\in \bigcup_{h=0}^{\ell-1}\mathcal{F}_{h}$ such that
$S=S'-S''$,
the unit-vector $\vec{v}_{\{i\}}$ is in $S'$, and
$S''$ contains a unit-vector $\vec{v}_{\{j\}}$ with $j<i$.
(Possibly, $S'\in\F_{\ell}$, i.e., $S'$ could be in the same sub-family as $S$.
Note that the induction is valid, because $S=S'-S''$ and $S',S''$ cross, hence, $|S'|>|S|$.)

\item{Induction Basis:}
This applies to the sub-family $\F_1$, with $\ell=1$.
The sets added to $\F_1$ by the algorithm have the form $V_i-V_j$ or $V_i\cap{V_j}$ for indices $i,j\in[k], i\not=j$.
Observe that each unit-vector is in exactly one of the sets $V_1,\dots,V_j$, hence,
any set of the form $V_i\cap{V_j}$ has no unit-vectors.
Then, by the construction used in the algorithm, $S = V_i-V_j$ for indices $i,j\in[k], j<i$.
Thus, the induction basis holds.

\item{Induction Step:}
Let $S \in \F_{\ell + 1}$ be a set that contains a unit-vector, say $\vec{v}_{\{i\}}\in{S}$.
Since Algorithm~\ref{alg:csccounter} added $S$ to $\F_{\ell+1}$,
there is a pair of crossing sets $A,B\in\bigcup_{h = 0}^\ell \F_h$ such that
$S=A\cap{B}$ or $S=A-B$ or $S=B-A$ (and the algorithm added $S$ to $\F_{\ell+1}$ due to $A,B$).
\\
\begin{tabularx}{0.8\textwidth}{cX}
$\circledast$ \quad &
If $S$ is a set difference of $A,B$, then we fix the notation such that $S=A-B$, and
if $S=A\cap{B}$, then we pick $A,B$ such that $|A| \geq |B|$ and $|A|$ is as large as possible
(among all crossing pairs of sets $A,B\in\bigcup_{h = 0}^\ell \F_h$ such that $S=A\cap{B}$).
\end{tabularx}

\item{Case 1:}
Suppose $S=A-B$; note that $\uv{i}\in{S}$.
By Step~(b)(i) of Algorithm~\ref{alg:csccounter}, $A\owns\uv{i}$, and
$B$ contains a unit-vector $\uv{j}$ with $j<i$.
Thus $S'=A,S''=B$ and we are done.

\item{Case 2:} Now suppose $S = A \cap B$.
Since $\uv{i}\in{S}$, note that $A,B$ are proper subsets of $V_i$ 
(if either $A=V_i$ or $B=V_i$ then $A,B$ would not cross).

Since $A\notin\{V_1,\dots,V_k\}$, $A\owns\uv{i}$, and $A\in\bigcup_{h=0}^{\ell}\mathcal{F}_{h}$
(note that $S\in\F_{\ell+1}$, $A\notin\F_{\ell+1}$),
by the induction hypothesis,
there is a crossing pair of sets $A',A''$ such that
$A=A'-A''$,
$A' \in \bigcup_{h=0}^{\ell}\mathcal{F}_{h}$, $A'' \in \bigcup_{h=0}^{\ell-1}\mathcal{F}_{h}$,
$\uv{i}\in A'$, $\uv{j}\in A''$, and $j < i$.

Thus, we have $S = (A' - A'') \cap B$, and this is equivalent to $S = (A' \cap B) - A''$.

\item{Subcase 2.1:}
Suppose $A',B$ cross. Then the algorithm adds $A'\cap{B}$ to $\F$, and we have
$A'\cap{B}\in \bigcup_{h=0}^{\ell+1}\mathcal{F}_{h}$ since $A',B\in \bigcup_{h=0}^{\ell}\mathcal{F}_{h}$.  
Recall that $\uv{i}\in{A\cap{B}}\subseteq{A'\cap{B}}$ and $\uv{j}\in{A''}$.
Next, observe that $A' \cap B \cap A''$ is non-empty, hence, by Fact~\ref{fact:crossing},
$A' \cap B$ and $A''$ cross.
(If $A' \cap B \cap A''$ is empty, then we would have $S = (A' \cap B) - A'' = (A' \cap B)$, and
this would contradict our choice of $A,B$ since $A' \supsetneq A$, and,
by $\circledast$, we would choose $A',B$.)
Thus $S'=A'\cap{B},S''=A''$ and we are done.

\item{Subcase 2.2:}
Suppose $A'$ and $B$ do not cross.
First, note that $B$ is a proper subset of $A'$,
because $A'\cap{B}$ is non-empty ($\uv{i}\in A'\cap{B}$) and $A-B \subset A'$ (since $A=A'-A''$).
Next, observe that $B\cap{A''}$ is non-empty;
otherwise, if $B\cap{A''}$ is empty, we would have a contradiction: $S=(A'\cap{B})-A''=B-A''=B$.
Hence, $A''$ and $B$ cross because $\uv{i}\in{A\cap{B}}\subseteq{B}$ and $\uv{j}\in{A''}$
(apply Fact~\ref{fact:crossing}).
Finally, note that $S=(A'\cap{B})-A''=B-A''$, thus taking $S'=B,S''=A''$ we are done.
\end{description}
\end{proof}

\IGNORE{
Figure~\ref{fig:csccountersubcases} visualizes these subcases.

\begin{figure}
    \centering
    \label{fig:csccountersubcases}
\caption{Visualization of the subcases of Lemma~\ref{lem:keyclaim}, Case 2.}
\end{figure}
endIGNORE}

\begin{lemma}\label{lem:rewriteSinF}
Let $S\in\F_{\ell}, \ell\geq1$, be a set that contains a unit-vector, say $\vec{v}_i \in S$, $i\in[k]$.
Then $S$ can be written as an expression, denoted express$(S,i)$, in terms of the sets $V_1,\dots,V_i$
(i.e., the sets of $\F_0$ with index in $[i]$) such that
express$(S,i)$ has the form $\big($express$(S',i)$ $-$ express$(S'',\hat{j})\big)$ where $\hat{j}<i$.
Moreover, the first term in express$(S,i)$ is $V_i$
and every other (``bottom~level'') term in this expression has the form $V_{j}, j<i$.
\end{lemma}
\begin{proof}
We repeatedly apply Lemma~\ref{lem:keyclaim}, starting with the expression $S=S'-S''$, where
	$S'\owns\vec{v}_i$, $S'\in  \bigcup_{h=0}^{\ell}\mathcal{F}_{h}$ and
	$S''\owns\vec{v}_{\hat{j}}$, ${\hat{j}}<i$, $S''\in \bigcup_{h=0}^{\ell-1}\mathcal{F}_{h}$,
until each set $R$ in express$(S,i)$ is a set of $\F_0$ (i.e., $R\in\{V_1,\dots,V_k\}$).

Whenever we apply Lemma~\ref{lem:keyclaim} to rewrite a set $R$ in the form $R'-R''$,
note that $R''$ is from an earlier iteration than $R$ (i.e., $R''\in\F_{\ell''}$ where $\ell''<\ell$), and
$R'$ is either from an an earlier iteration than $R$ or
it is from the same iteration as $R$, and, in the latter case, we have $|R'| > |R|$
(because $R',R''$ is a crossing pair of sets such that $R=R'-R''$).
Let us denote the unit-vector in $R$ by $\uv{i'}$
(thus, $R\subset{V_{i'}}$, $R\not=V_{i'}$).
Note that $R'$ contains the unit-vector $\uv{i'}$ (this is the unit-vector in $R$)
and $R''$ contains a unit-vector $\uv{j'}$, where $j'<i'$.
Therefore, the rewriting process terminates with an expression in terms of the sets $V_1,\dots,V_i$.

Moreover, observe that the first term in the expression express$(S,i)$ is $V_i$
and every other ``bottom~level'' term in this expression has index less than $i$.
In more detail, if we represent the parenthesized expression express$(S,i)$ as a binary tree that has
a node representing each set $R$ that is rewritten in the form $R'-R''$ via Lemma~\ref{lem:keyclaim},
then, the bottom~level nodes of this tree represent the sets $V_1,\dots,V_k$,
the first (left~most) bottom~level node represents $V_i$, and
each of the other bottom~level nodes represents one of the sets $V_1,\dots,V_{i-1}$.
\end{proof}

\begin{lemma}\label{lem:propertySinF}
Let $S\in\F_{\ell}, \ell\geq1$, be a set that contains a unit-vector, say $\vec{v}_i \in S$, $i\in[k]$.
Let $I$ be the index set $\{i\} \cup I_{\oplus}$ where $I_{\oplus}$ is a subset of $\{i+1,\dots,k\}$.
Then $S$ contains the vector $\vec{v}_I$. Therefore, $|S| \geq 2^{k-i}$.
\end{lemma}
\begin{proof}
By Lemma~\ref{lem:rewriteSinF}, we can rewrite $S$ in terms of of the sets $V_1,\dots,V_i$
in the form express$(S,i)$ such that
express$(S,i)$ has the form $\big($express$(S',i)$ $-$ express$(S'',\hat{j})\big)$ where $\hat{j}<i$.
Moreover, the first term in express$(S,i)$ is $V_i$
and every other term in this expression has the form $V_{j}, j<i$.

Clearly, $V_i$ contains the vector $\vec{v}_I$, and, moreover,
$\vec{v}_I$ is in none of the sets $V_{j}, j<i$
(note that every vector $\vec{v}\in{V_{j}}$ has $\vec{v}_{j}=1$,
whereas the vector $\vec{v}_I$ has a zero in the $j$-th coordinate).
Hence, by the properties of express$(S,i)$, $S$ contains $\vec{v}_I$.

Observe that there are $2^{k-i}$ index sets of the form $I$ (since there are
$2^{k-i}$ distinct subsets of $\{i+1,\dots,k\}$).
\end{proof}

\begin{lemma}\label{lem:largecuts}
The family of sets $\F$ computed by Algorithm~\ref{alg:csccounter} satisfies the following:
\begin{enumerate}[(a)]
    \item For each $i \in [k-1]$, $\{ \uv{i} \}\notin \F$.
    \item For $i,j\in[k]$ with $i<j$, $V_i-V_j\notin \F$.
    \item For $i\in\{3,\dots,k\}$, let $W_i$ be a set such that
	$\uv{i}\in{W_i}$, $\vec{v}_{[k]}\in{W_i}$, and $\vec{v}_{\{1,i\}}\notin{W_i}$. Then $W_i\notin\F$.
\end{enumerate}
\end{lemma}
\begin{proof} 
\begin{description}
\item[Part~(a):]
By Lemma~\ref{lem:propertySinF}, any set $S\in\F$ that contains a unit-vector $\uv{i}$, $i\in[k-1]$,
has size $\geq2^{k-i} \geq 2$.
Hence, for $i \in [k-1]$, $\F$ does not contain the singleton-set containing the unit-vector $\uv{i}$.

\item[Part~(b):]
Observe that the vector $\vec{v}_{\{i,j\}}$ is in both $V_i$ and $V_j$, so it is not in the set $V_i-V_j$.
On the other hand, by Lemma~\ref{lem:propertySinF}, if a set $S\in\F$ contains the unit-vector $\uv{i}$,
then $S$ also contains the vector $\vec{v}_{\{i,j\}}$.
Therefore, the set $V_i-V_j$ is not in $\F$.

\item[Part~(c):]
Observe that the vector $\vec{v}_{\{1,i\}}$ is in $V_1$ and $V_i$,
and it is in none of the sets $V_j$, $j\in\{2,\dots,k\}-\{i\}$.
By way of contradiction, suppose that $W_i$ is in $\F$;
note that $W_i\not=V_i$ (since $\vec{v}_{\{1,i\}}\in{V_i}$ and $\vec{v}_{\{1,i\}}\not\in{W_i}$).

By Lemma~\ref{lem:rewriteSinF}, we can rewrite $W_i$ in terms of of the sets $V_1,\dots,V_i$
in the form express$(W_i,i)$ such that
express$(W_i,i)$ has the form $\big($express$(S',i)$ $-$ express$(S'',\hat{j})\big)$ where $\hat{j}<i$.
Moreover, the first term in express$(W_i,i)$ is $V_i$
and every other term in this expression has the form $V_{j}, j<i$.

Since $\vec{v}_{\{1,i\}}\notin{W_i}$, it follows that the term $V_1$ occurs in express$(W_i,i)$,
that is, we are removing $V_1$ or a subset of $V_1$ from $V_i$ to obtain $W_i$.
Since $\vec{v}_{[k]}\in{W_i}$, we are removing a proper subset of $V_1$ from $V_i$
(otherwise, if we remove $V_1$ from $V_i$, then we would remove $\vec{v}_{[k]}$ from $W_i$).
We have a contradiction, since express$(W_i,i)$ has no sub-expression of the form $(V_1 - (V_j \dots))$
(because we would have $j<1$, by the definition of express$(W_i,i)$).
\end{description}
\end{proof}

\begin{proposition}\label{pro:partition-uncrossable}
The family of sets $\mathcal{F}$ cannot be partitioned into $d < k$ uncrossable families.
\end{proposition}
\begin{proof}
For $i,j\in[k]$, with $i<j$, observe that:
\begin{itemize}
\item[(a)]
$V_i \cup V_j$ is not in $\F$, because, by Lemma~\ref{lem:subsets},
every set in $\mathcal{F}$ is a subset of one of the sets $V_1,\dots,V_k$.
\item[(b)]
$V_i-V_j$ is not in $\F$, by Lemma~\ref{lem:largecuts}, part~(b).
\end{itemize}

Now, suppose that $\F$ could be partitioned into $d<k$ uncrossable families.
Then two of the sets $V_i$ and $V_j$, where $i,j\in[k], i<j,$ would
be in the same ``block'' of the partition, i.e.,
$V_i$ and $V_j$ would be in the same uncrossable family, call it $\hat{\F}$.
This would violate the uncrossability property, since $V_i\cup{V_j}\not\in\hat{\F}$ and
$V_i-V_j\not\in\hat{\F}$.
\end{proof}

}

\section{Symmetric submodular functions versus $\F$ \label{sec:submodular}}
{
In this section, our goal is to prove the following result.

\begin{proposition}\label{pro:no-submod-realization}
There do not exist a symmetric submodular function $g:2^V\rightarrow\Q$ and $\lambda\in\Q$
such that $\F = \{ S \,:\, g(S)<\lambda \}$.
\end{proposition}

Given a graph $G=(V,E)$ and non-negative capacities on the edges, $c:E\rightarrow \mathbb{Q}$,
the cut-capacity function $c(\delta_G(\cdot)):2^V\rightarrow\Q$ is a symmetric submodular function.
(Recall that $c(\delta_G(S)):=\sum_{e\in\delta_G(S)}c_e$.)
Proposition~\ref{pro:no-submod-realization} implies that the family $\F$ cannot be realized
as the family of small cuts of a capacitated graph;
in other words, there do not exist any capacitated graph $G=(V,E),c$ and $\lambda\in\Q$
such that $\F = \{S \,:\, c(\delta_G(S))<\lambda\}$.

We prove Proposition~\ref{pro:no-submod-realization} using the following contradiction argument.
Let $g(\cdot)$ be any symmetric submodular function on the ground~set $V$
(recall that $\F$ is a family of subsets of $V$).
For any pair of crossing sets $A,B\subseteq{V}$, we have the inequality
$g(A) + g(B) \geq g(A-B) + g(B-A)$.
Let $k\geq3$ be a positive integer.
Recall that Algorithm~\ref{alg:csccounter} starts with the sets $V_1,\dots,V_k$ and constructs $\F$.
Suppose there exist $g(\cdot)$ and $\lambda\in\Q$ such that $\F = \{S\,:\,g(S)<\lambda\}$.
We focus on $2k-3$ pairs of crossing sets (to be discussed below) and
the corresponding $2k-3$ inequalities.
Summing up these $2k-3$ inequalities, we obtain the inequality
\[ g(V_1) + g(V_2) + \dots + g(V_k) - g(\{\uv{1}\}) - g(\{\uv{2}\}) - g(W_3) - \dots - g(W_k) \geq 0, \quad(*)\]
where $W_3,\dots,W_k$ are subsets of $V$ that are not present in $\F$;
we will define $W_3,\dots,W_k$ in what follows.
Recall that $V_1,\dots,V_k\in\F$ and, by Lemma~\ref{lem:largecuts}(a), $\{\uv{1}\},\{\uv{2}\}\not\in\F$.
Since $V_1,\dots,V_k\in\F$, we have $g(V_i)<\lambda, \forall{i}\in[k]$.
Since $\{\uv{1}\},\{\uv{2}\}\not\in\F$ and $W_3,\dots,W_k\notin\F$, we have
$g(\{\uv{1}\})\geq\lambda, g(\{\uv{2}\})\geq\lambda, g(W_i) \geq\lambda, \forall{i}\in\{3,\dots,k\}$.
Hence, inequality~$(*)$ cannot hold. This gives the required contradiction.

Next, we list the $2k-3$ pairs of crossing sets, and, below, we illustrate inequality~$(*)$ for $k=4$.
The $2k-3$ pairs of crossing sets consist of two lists.
The first list has the following $k-1$ pairs of sets;
Lemma~\ref{lem:crossingpairs} (given below) shows that each of these is a pair of crossing sets:
\begin{align*}
V_1,& \quad V_2 \\
(V_1-V_2),& \quad V_3 \\
(V_1-V_2-V_3),& \quad V_4 \\
\dots \\
(V_1-V_2-\dots-V_{k-1}),& \quad V_k.\\
\end{align*}
For $i=3,\dots,k$, we define $U_i$ to be the set
$V_i - (V_1 - V_2 - \dots - V_{i-1})$.
Thus, $U_3=V_3 - (V_1-V_2)$, $U_4=V_4 - (V_1-V_2-V_3)$, \dots, $U_k=V_k - (V_1-V_2-\dots-V_{k-1})$.
For $i=2,\dots,k-1$, note that $U_{i+1}$ is one of the set~differences for
the $i$-th pair of crossing sets listed above.
The second list has the following $k-2$ pairs of sets;
Lemma~\ref{lem:crossingpairs} (given below) shows that each of these is a pair of crossing sets:
\begin{align*}
(V_2-V_1),& \quad U_3 \\
(V_2-V_1-V_3),& \quad U_4 \\
\dots \\
(V_2-V_1-V_3-\dots-V_{k-1}),& \quad U_k.\\
\end{align*}
We define $W_3$ to be the set $U_3 - (V_2-V_1)$, and
for $i=4,\dots,k$, we define $W_i$ to be the set
$U_i - ((V_2 - V_1) - V_3 -\dots- V_{i-1})$.
Thus, $W_3=U_3 - (V_2-V_1)$, $W_4=U_4 - (V_2-V_1-V_3)$, \dots, $W_k=U_k - (V_2-V_1-\dots-V_{k-1})$.
Note that the two set~differences for
the first pair of crossing sets in the second list above are
$(V_2-V_1) - U_3$ and $W_3$, and
for $i=2,\dots,k-2$, the two set~differences for
the $i$-th pair of crossing sets in the second list above are
$((V_2-V_1)-V_3-\dots-V_{i+1}) - U_{i+2}$ and $W_{i+2}$.
Moreover, note that
$(V_2-V_1) - U_3 = (V_2-V_1) - \big(V_3-(V_1-V_2)\big) = (V_2-V_1-V_3)$, and
for $i=2,\dots,k-2$, note that
$((V_2-V_1)-V_3-\dots-V_{i+1}) - U_{i+2} =
((V_2-V_1)-V_3-\dots-V_{i+1}) - \big(V_{i+2} - (V_1-V_2-\dots-V_{i+1})\big) =
((V_2-V_1)-V_3-\dots-V_{i+2})$, because the set of the first term, $((V_2-V_1)-V_3-\dots-V_{i+1})$,
is disjoint from the set $(V_1-V_2-\dots-V_{i+1})$.

\begin{lemma}\label{lem:crossingpairs}
\begin{enumerate}[(a)]
\item In the first list, every pair of sets is crossing.
\item In the second list, every pair of sets is crossing.
\end{enumerate}
\end{lemma}
\begin{proof}
\begin{description}
\item[(a):]
Let $i$ be an index in $\{1,\dots,k-1\}$.
The $i$-th pair of sets in the first list is
$(V_1-V_2-\dots-V_i), V_{i+1}$.
Note that the unit-vector $\uv{1}$ is in the set $(V_1-V_2-\dots-V_i)$,
and the unit-vector $\uv{i+1}$ is in the set $(V_{i+1})$.
The intersection of the two sets is non-empty, since the vector $\vec{v}_{\{1,i+1\}}$ is in
both sets.
Then, by Fact~\ref{fact:crossing}, the two sets are crossing.

\item[(b):]
Let $i$ be an index in $\{1,\dots,k-2\}$.
The first pair of sets in the second list is
$(V_2-V_1), U_{3}$.
For $i\geq2$, the $i$-th pair of sets in the second list is
$((V_2-V_1)-V_3-\dots-V_{i+1}), U_{i+2}$.
Note that the unit-vector $\uv{2}$ is in the first set
(namely, $(V_2-V_1)$ or $((V_2-V_1)-V_3-\dots-V_{i+1})$),
and the unit-vector $\uv{i+2}$ is in the second set (namely, $(U_{i+2})$).
The intersection of the two sets is non-empty, since the vector $\vec{v}_{\{2,i+2\}}$ is in
both sets.
Then, by Fact~\ref{fact:crossing}, the two sets are crossing.
\end{description}
\end{proof}

\begin{lemma}\label{lem:W-sets}
The set $W_3 = U_3 - (V_2-V_1)$ is not present in $\F$, and
for $i=4,\dots,k$, the set $W_i = U_i - ((V_2 - V_1) - V_3 -\dots- V_{i-1})$ is not present in $\F$.
\end{lemma}
\begin{proof}
Let $i$ be an index in $\{4,\dots,k\}$.
Observe that
$W_i = U_i - ((V_2 - V_1) - V_3 -\dots- V_{i-1}) =
V_i - (V_1 - V_2 - V_3 -\dots- V_{i-1}) - ((V_2 - V_1) - V_3 -\dots- V_{i-1})$.
Clearly, the unit-vector $\uv{i}$ is in $W_i$, since
$\uv{i}\in V_i$ and, for $j=1,\dots,i-1$, $\uv{i}\notin V_j$.
Moreover, the vector $\vec{v}_{[k]}$ is in $W_i$, since this vector is in $V_i$
and this vector is not in either of the sets
$(V_1 - V_2 - V_3 -\dots- V_{i-1})$ or $((V_2 - V_1) - V_3 -\dots- V_{i-1})$.
The vector $\vec{v}_{\{1,i\}}$ is not in $W_i$,
since this vector is in the sets $V_1, V_i$ and, for $j=1,\dots,i-1$, $\vec{v}_{\{1,i\}}\notin V_j$,
hence, $\vec{v}_{\{1,i\}}$ is in both the sets $V_i$ and $(V_1 - V_2 - V_3 -\dots- V_{i-1})$.
Then, by Lemma~\ref{lem:largecuts}(c), the set $W_i$ is not in $\F$.

Similar arguments show that $W_3\notin\F$.
\end{proof}

When we sum the $2k-3$ inequalities corresponding to the $2k-3$ pairs of crossing sets,
then several of the terms cancel out, leaving only the terms
$g(V_1),\dots,g(V_k)$, $-g(\{\uv{1}\})$, $-g(\{\uv{2}\})$, $-g(W_3),\dots,-g(W_k)$.
In more detail, for the $i$-th crossing pair in the first list,
for $i\in\{1,\dots,k-2\}$,
the term $-g(V_1-V_2-\dots-V_{i+1})$ cancels with the term $+g(V_1-V_2-\dots-V_{i+1})$
of the $(i+1)$-th crossing pair in the first list, and
for $i\in\{2,\dots,k-1\}$, the term $-g(U_{i+1})$ cancels with
a term of the $(i-1)$-th crossing pair in the second list;
the term $-g(V_2-V_1)$ of the first crossing pair in the first list cancels with
a term of the first crossing pair in the second list.
Lastly, for the $i$-th crossing pair in the second list,
for $i\in\{1,\dots,k-3\}$,
the term $-g((V_2-V_1)-V_{3}-\dots-V_{i+2})$ cancels with the term $+g((V_2-V_1)-V_{3}-\dots-V_{i+2})$
of the $(i+1)$-th crossing pair in the second list.

Proposition~\ref{pro:no-submod-realization} follows from the above
discussion and Lemmas~\ref{lem:crossingpairs}, \ref{lem:W-sets}.
The next result follows from the propositions.

\begin{theorem}\label{thm:main}
For any positive integer $d\geq2$, Algorithm~\ref{alg:csccounter}
constructs a pliable family of sets $\F$ that satisfies structural submodularity such that
(a)~there do not exist a symmetric submodular function $g:2^V\rightarrow\Q$ and $\lambda\in\Q$ such that
\hbox{$\F = \{ S \,:\, g(S)<\lambda \}$}, and
(b)~$\F$ cannot be partitioned into $d$ (or fewer) uncrossable families.
\end{theorem}

\bigskip
\bigskip

\hbox{
\begin{minipage}{\textwidth}
{
\begin{example}
The following example with $k=4$ illustrates the above discussion.

Note that $U_3 = V_3 - (V_1-V_2)$, $U_4=V_4 - (V_1-V_2-V_3)$, and
$W_3 = U_3 - (V_2-V_1)$, $W_4 = U_4 - (V_2-V_1-V3)$.
Also, note that $(V_1-V_2-V_3-V_4) = \{\uv{1}\}$, and $(V_2-V_1-V_3-V_4) = \{\uv{2}\}$.

We have $2k-3 = 5$ pairs of crossing sets, and the corresponding inequalities.

\begin{align*}
V_1,& \quad V_2 		& g(V_1) + g(V_2) - g(V_1-V_2) - g(V_2-V_1) &\geq 0 \\
(V_1-V_2),& \quad V_3		& g(V_1-V_2) + g(V_3) - g(V_1-V_2-V_3) - g(U_3) &\geq 0 \\
(V_1-V_2-V_3),& \quad V_4	& g(V_1-V_2-V_3) + g(V_4) - g(V_1-V_2-V_3-V_4) - g(U_4) &\geq 0 \\
(V_2-V_1),& \quad U_3		& g(V_2-V_1) + g(U_3) - g(V_2-V_1-V_3) - g(W_3) &\geq 0 \\
(V_2-V_1-V_3),& \quad U_4	& g(V_2-V_1-V_3) + g(U_4) - g(V_2-V_1-V_3-V_4) - g(W_4) &\geq 0 \\
\text{Sum of inequalities:}	&&
	g(V_1) + g(V_2) + g(V_3) + g(V_4) - g(\{\uv{1}\}) - g(\{\uv{2}\}) - g(W_3) - g(W_4) &\geq 0
\end{align*}
\end{example}
}
\end{minipage}
}
}
\bibliographystyle{plainurl}
\bibliography{sbc-csc-submod}

@article{BCGI24,
  author = {Ishan Bansal and Joseph Cheriyan and Logan Grout and Sharat Ibrahimpur},
  title = {Improved Approximation Algorithms by Generalizing the Primal-Dual Method Beyond Uncrossable Functions},
  journal = {Algorithmica},
  volume = {86},
  number = {8},
  pages = {2575--2604},
  year = {2024},
  url = {https://doi.org/10.1007/s00453-024-01235-2},
  doi = {10.1007/s00453-024-01235-2},
  timestamp = {Tue, 30 Jul 2024 01:00:00 +0200},
  biburl = {https://dblp.org/rec/journals/algorithmica/BansalCGI24.bib},
  bibsource = {dblp computer science bibliography, https://dblp.org},
  _bib2doi_selected = {dblp:/rec/journals/algorithmica/BansalCGI24.bib},
  _bib2doi_confirmed = {true},
  _bib2doi_finished = {true},
}

@article{WGMV95,
  author = {David P. Williamson and Michel X. Goemans and Milena Mihail and Vijay V. Vazirani},
  title = {A Primal-Dual Approximation Algorithm for Generalized {S}teiner Network Problems},
  journal = {Combinatorica},
  volume = {15},
  number = {3},
  pages = {435--454},
  year = {1995},
  doi = {10.1007/BF01299747},
  timestamp = {Wed, 22 Jul 2020 01:00:00 +0200},
  biburl = {https://dblp.org/rec/journals/combinatorica/WilliamsonGMV95.bib},
  bibsource = {dblp computer science bibliography, https://dblp.org},
  _bib2doi_selected = {dblp:/rec/journals/combinatorica/WilliamsonGMV95.bib},
  _bib2doi_confirmed = {true},
  _bib2doi_finished = {true},
}

@article{D2017,
   title={Abstract Separation Systems},
   volume={35},
   ISSN={1572-9273},
   url={http://dx.doi.org/10.1007/s11083-017-9424-5},
   DOI={10.1007/s11083-017-9424-5},
   number={1},
   journal={Order},
   publisher={Springer Science and Business Media LLC},
   author={Diestel, Reinhard},
   year={2017},
   month=apr, pages={157–170} }

@article{DEW2019,
title = {Structural submodularity and tangles in abstract separation systems},
journal = {Journal of Combinatorial Theory, Series A},
volume = {167},
pages = {155-180},
year = {2019},
issn = {0097-3165},
url = {https://doi.org/10.1016/j.jcta.2019.05.001},
author = {Reinhard Diestel and Joshua Erde and Daniel Weißauer},
}

@misc{EKT2021,
      title={The Structure of Submodular Separation Systems}, 
      author={Christian Elbracht and Jay Lilian Kneip and Maximilian Teegen},
      year={2021},
      eprint={2103.13162},
      archivePrefix={arXiv},
      primaryClass={math.CO},
      url={https://arxiv.org/abs/2103.13162}, 
}

@article{DEE2017,
author = {Diestel, Reinhard and Eberenz, Philipp and Erde, Joshua},
title = {Duality Theorems for Blocks and Tangles in Graphs},
journal = {SIAM Journal on Discrete Mathematics},
volume = {31},
number = {3},
pages = {1514-1528},
year = {2017},
doi = {10.1137/16M1077763},
URL = {https://doi.org/10.1137/16M1077763},
}

@article{DO2019,
author = {Diestel, Reinhard and Oum, Sang-il},
title = {Tangle-Tree Duality: In Graphs, Matroids and Beyond},
year = {2019},
issue_date = {Aug 2019},
publisher = {Springer-Verlag},
address = {Berlin, Heidelberg},
volume = {39},
number = {4},
issn = {0209-9683},
url = {https://doi.org/10.1007/s00493-019-3798-5},
doi = {10.1007/s00493-019-3798-5},
journal = {Combinatorica},
month = aug,
pages = {879–910},
numpages = {32}
}

@mastersthesis{S2025,
  title        = {Cover {S}mall {C}uts and {F}lexible {G}raph {C}onnectivity {P}roblems},
  author       = {Miles Simmons},
  year         = 2025,
  month        = {September},
  address      = {Waterloo, ON, Canada},
  url          = {https://hdl.handle.net/10012/22426},
  school       = {University of Waterloo},
  type         = {Master's thesis}
}

@book{Schrijver,
  title     = {{Combinatorial Optimization: Polyhedra and Efficiency}},
  author    = {Alexander Schrijver},
  volume    = {24},
  series    = {Algorithms and Combinatorics},
  year      = {2003},
  publisher = {Springer},
  address = {Berlin Heidelberg New York},
}

\end{document}